\documentclass[twocolumn,showpacs,preprintnumbers,amsmath,amssymb]{revtex4}

\usepackage{graphicx}
\usepackage{dcolumn}
\usepackage{bm}
\usepackage[dvips]{epsfig}

\begin{document}

\title{The S shape of a granular pile in a rotating drum}

\author{Nicolas Taberlet$^{1,2}$}
\author{Patrick Richard$^1$}
\author{E. John Hinch$^2$}

\affiliation{1. GMCM, Universit\'e Rennes 1, CNRS UMR 6626, Batiment 11A, 35042 Rennes, France\\
2. DAMTP, University of Cambridge, Wilberforce Road, Cambridge, CB3 0WA, 
United Kingdom}

\date{\today}

\begin{abstract}
The shape of a granular pile in a rotating drum is investigated. Using Discrete Elements Method (DEM) simulations we show that the ``S shape" obtained for high rotation speed can be accounted for by the friction on the end plates. A theoretical model which accounts for the effect of the end plates is presented and the equation of the shape of the free surface is derived. The model reveals a dimensionless number which quantifies the influence of the end plates on the shape of the pile. Finally, the scaling laws of the system are discussed and numerical results support our conclusions.
\end{abstract}


\maketitle
Among all the geometries used to study granular flows, the rotating drum
might be the most complex~\cite{Henein1983b,Mellmann2001,gdrmidi04}.  
Depending on the angular velocities, two different regimes occur.
At low rotation speed the free surface of the pile is inclined and flat, but
a significant curvature appears at high rotation speed: the so called S shape~\cite{Rajchenbach1990,nakagawa99,Orpe2001}.
Up until now this transition remains relatively unexplored~\cite{Henein1983b,Felix2002}. 
Different explanations have been proposed, but no consensus has been reached.
One possible origin of the curvature of the bed is the centrifugal forces acting on the 
granular flow. This effect can be quantified by the Froude number: $Fr= R \, \Omega^2 /g$, 
where $\Omega$ is the angular rotation speed of the drum, 
$R$ its radius and $g$ the gravitational acceleration.
{Another possible cause of the S shape is the ``feeding inertia" of the grains. 
If the rotation speed is high then the velocity of the grains, $v$, can be very high and the grains can display ballistic trajectories. The vertical distance, $\Delta$, traveled  by such a grain should be compared to the radius of the drum. This effect is again quantified by the Froude number: $\Delta/R = v^2/(2 g d) = R^2  \Omega^2 / (2 g R) = Fr/2$. 
Although these effects can indeed account for the curvature of the free surface, the S shape can be observed for very low values of the Froude number (typically, $Fr=10^{-4}$, see~\cite{Felix2002} or $Fr=10^{-3}$ in fig. 2). This indicates that, in this case, the centrifugal force is not the origin of the S shape. Another explanation is then needed.

Several other models have been proposed~\cite{Zik1994,Elperin1998,levine99,Mellmann2001} and were able to recover the S shape. Yet, none has taken into account the influence of the end plates. Here instead, we show that in short drums, the S shape can be explained by the friction of the end plates of the cylinders.} This idea is based on recent work on confined granular flows which have shown that the side-walls of a channel can drastically influence the flow
properties~\cite{khakhar01,Courrech03,taberlet03,caprihan05,renouf05,Jop05}. In particular, Khakhar {\it et 
al.}~\cite{khakhar01} reported that the inclination of a flow on a heap increases with increasing flow rate, a 
phenomenon  later explained by Taberlet {\it et al.}~\cite{taberlet03}. These authors have derived a law linking the 
inclination of the free surface, $\varphi$, with the flow thickness, $h$, and the channel width, $L$:
\begin{equation}
\tan \varphi = \mu_i + \mu_w \, \frac{h}{L} \, ,
\label{equ1}
\end{equation}
\noindent where $\mu_i$ and $\mu_w$ are two effective friction coefficients describing the internal and side-walls frictional properties, respectively (see~\cite{taberlet03} for details).
In a rotating drum, the flow rate is the highest near the center of the drum, which is also the point at which the free surface is the steepest. This supports the idea that the end plates can have a crucial influence on the shape of the free surface.
The outline of the paper is as follows: 
first the simulation method is presented. The crucial effect of the end plates on the shape of the pile is then evidenced by two numerical tests: one with frictional and one with frictionless end plates. A theoretical model is presented and the equation of the free surface is derived. Finally, the scaling laws of the system are discussed.

\begin{figure}[htbp]
\begin{center}
\resizebox{6.6cm}{!}{\includegraphics*{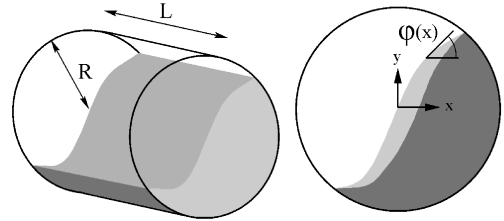}} \caption{
We consider a half-filled three dimensional
drum of radius $R$ and length $L$. The
local slope of the free surface is denoted $\varphi (x)$. Right: light
grey corresponds to flowing material and dark grey to solid rotation.
}\label{fig1}
\end{center}
\end{figure}

In this paper, we consider a cylindrical drum of radius $R$ which rotates at
a constant angular velocity $\Omega$. The drum is partially filled with
granular material. The position along the horizontal and vertical axes are labeled $x$ and $y$, respectively, and the local slope of the free surface is denoted $\varphi(x)$ (see Fig. 1).
For all the results shown in this paper, $\Omega$ is large enough to produce continuous
flows but small enough to neglect centrifugal forces and the feeding inertia effects, the Froude number being kept below 1.

The shape of the free surface of the pile is investigated through
numerical simulations of soft-sphere molecular dynamics
method. Although not flawless, this type of simulation has been widely
used in the past two decades and has proven to be very reliable for
the study of granular flows in a rotating
drum~\cite{Dury99,Taberlet04,taberlet05}. The forces acting between two colliding
grains are computed from the normal overlap, $\delta_n$, and the
equations of motion (displacement and rotation) are integrated using
the Verlet method~\cite{Frenkel96}. The schemes used for the forces
calculations are the spring-dashpot and the regularized Coulomb laws,
with the following values: particle diameter $d$=8 mm, mass=0.16 g,
spring constant $k_n$=40000 N.$\text m^{-1}$, viscous damping
$\gamma_n$=0.5 $\text s^{-1}$, leading to a normal coefficient of
restitution $e_n=0.64$, regularization constant $\gamma_t$=5 $\text s^{-1}$, time step $dt=10^{-6}$ s and, unless otherwise mentioned, $\mu$=0.3. Note that the restitution, i.e. the inelasticity, does not seem to play a role. Different values of $e_n$ were tried ($0.3<e_n<0.8$) and did not affect the shape of the pile.
The collisions against the wall and the end plates are treated like particle-particle collisions with one of the particles having infinite mass and radius, which mimics a flat surface.
Note that the mechanical properties of the grain/end plate collisions can be chosen independently of those of the grain/grain collisions. In particular, it is possible to simulate frictional grains placed in a drum with frictionless end plates.

\indent The radius of the drum $R$ is typically 100$d$, and its length, $L$, is varied from 10$d$ to 200$d$. The grains are first released in the drum and rotation is started (at
rotation speed $\Omega$) only after they have settled. The number of
grains is chosen so that the drum is half-filled. Our simulations
contain a large number of particles, between 5\,000 and 70\,000, and run
for typically 5 full rotations of the drum.
The granular material is made slightly polydisperse with an equal number of grains of diameter $d$ and 4/5$d$, in order to avoid crystallization.
The snapshots of the pile were taken after two full rotations after
the shape has reached a steady state. On such short times no radial (or axial) segregation was observed. {Note that no interstitial fluid is present in our simulation but its effect is expected to be negligible for the low values of $\Omega$ (typically $Fr=10^{-3}$.}

\begin{figure}[htbp]
\begin{center}
\resizebox{6.6cm}{!}{\includegraphics*{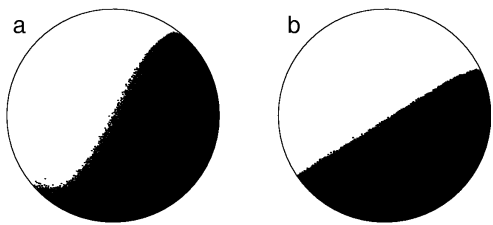}} \caption{
Side views of the 3D drum. In both cases, $R=80 \,d$, $L=10 \,d$, $\mu=0.3$, $\Omega$=0.1 rad/s and 
$N$=36\,000 grains. a) frictional and b) frictionless end plates. The difference in shape is very clear and originates from a change in the frictional properties of the end plates. This provides evidence of the crucial role of the end plates on the shape of the free surface.
}\label{fig2}
\end{center}
\end{figure}

In order to demonstrate the crucial influence of the end plates, and
in particular their frictional properties, two simulations differing
only by their end plate friction coefficients were performed. Figure 2
shows two runs with identical values of all parameters, one with
frictional end plates (a) and the other with frictionless end plates (b). The two
shapes are very different, displaying on one hand the S shape (a), and on the other hand 
a flat surface (b). The rotation speed
being the same in both cases (meaning that $Fr$ is the same) 
this result gives strong evidence that the end plates have
a crucial influence on the shape of the free surface and shows that
neither the centifugal force nor the feeding inertia is responsible for the S shape.

Let us now present a theoretical model based on Eq.(1). 
The aim of our model is to derive the equation for the position of the free surface, that is, to find an expression for $y_{surf}(x)$. The following notations will be used:
$\bar{\rho}$ is the average density of the material (in kg\,m$^{-3}$),
$q_{feed}(x)$ is the local feeding rate per unit length and width along 
the flow (in kg\,s$^{-1}$m$^{-2}$),
$Q_{flow}(x)$ is the local flow rate per unit width (in kg\,s$^{-1}$m$^{-1}$),
$\dot{\gamma}$ is the shear rate,
$h(x)$ is the local thickness of the flow,
$g$ is the gravity,
and $d$ is the grain diameter.

Some authors mentioned the dependence of $\dot\gamma$
on the flow properties~\cite{Felix2002,rajchenbach00}.
Yet, in many cases this dependence is weak~\cite{bonamy02} so, in order to
simplify our analysis we assume that the shear rate is a constant.
For geometrical reasons, the feeding rate, $q_{feed}$, increases
linearly with the distance from the center of the drum, $r$, 
and is positive in one half of the drum ($x>0$) and negative in the other half.
The feeding rate then reads

\begin{equation}
q_{feed}(x) = \bar{\rho} \; \Omega \; r. \, 
\label{equ2}
\end{equation}

\noindent Because of mass conservation, the flow rate at a given point of the
free surface (at the postion \{$x$, $y_{surf}(x)$\}) has to be equal to
the integral of the feeding rate coming from the solid rotation
between the outer cylinder and the considered radius, 
$r=\sqrt{x^2+y_{surf}^2}$ :


\begin{equation}
\begin{array}{crl}
\displaystyle Q_{flow}(x) &=& \displaystyle \int_r^R \, q_{feed}(r) \, dr
=\displaystyle \frac{ \bar{\rho} \;  \Omega}{2} \, (R^2 - r^2),\\[3mm]
&=&\displaystyle \frac{ \bar{\rho} \;  \Omega}{2} \, (R^2 - x^2 - y_{surf}^2).
\label{equ3}
\end{array}
\end{equation}

\noindent To go further, one needs a relation between the flow thickness and flow rate.
{Previous theories for the shape, e.g.,~\cite{Zik1994,levine99}, have made different assumptions about the velocity profiles and the thickness of the flowing layer but those predate experimental observations. Here we build a model based on the observations in~\cite{gdrmidi04} which report a linear velocity profile with a near universal shear rate.}
Since we consider that the shear rate is a constant, dimensional
analysis yields
$\dot\gamma= a \sqrt{g/d}$. For simplicity we use $a=1$ but this value has no effect on the qualitative results presented below. Hence the mass flow in the layer of thickness $h$ with the linear velocity gradient $\dot{\gamma}$ is

\begin{equation}
Q_{flow}(x) = \frac{1}{2} \, \bar{\rho} \, \dot{\gamma} \, h(x)^2 = \displaystyle \frac{\bar{\rho}}{2} \, \sqrt{\frac{g}{d}} \, h(x)^2.
\label{equ4}
\end{equation}

\noindent The slope of the free surface (i.e. $\partial y_{surf}/\partial x$) is given by equation (1)

\begin{equation}
\displaystyle \frac{\partial y_{surf}}{\partial x} = \tan \varphi(x) = \displaystyle  \mu_i + \mu_w \, \frac{h(x)}{L}.
\label{equ5}
\end{equation}

\noindent Using (3) and (4), one can express $h$ as a function of $x$ and plugging it into (5) leads to:

\begin{equation}
\displaystyle \frac{\partial y_{surf}}{\partial x} =  \displaystyle
\mu_i + \mu_w \, \left(\frac{d \, \Omega^2}{g}\right)^{1/4}  \,   \frac{1}{L} \,  \sqrt{R^2-x^2-y_{surf}^2}.
\label{equ6}
\end{equation}



Equation (6) shows that the slope has a minimum value
$\mu_i$ at the outer cylinder and a maximum at the center.
It can be numerically integrated and the shape of the pile can be plotted for different values of the parameters (see Fig. 3).
{ The equation is integrated outwards starting from the center of the drum ($x=y=0$) using the Euler method. Note that no additional condition is needed since the size of the drum is embedded in Eq. (6) while its symmetry ensures a total mass of 1/2.}
Note, however, that Eqs.(1)-(6) do not require any assumptions regarding the filling ratio. Therefore, equation (6) can be integrated from any point, for instance on the outer boundary, after what the corresponding filling ratio could be computed.
For frictionless end plates (i.e. $\mu_w$=0), the free surface is a flat plane of equation $y= \mu_i \,x$, which confirms the results of Fig. 2. With increasing influence of the end plates a curvature appears and the pile displays an S shape, which qualitatively reproduces the experimental and numerical behavior. 

\begin{figure}[htbp]
\begin{center}
\resizebox{5.7cm}{!}{\includegraphics*{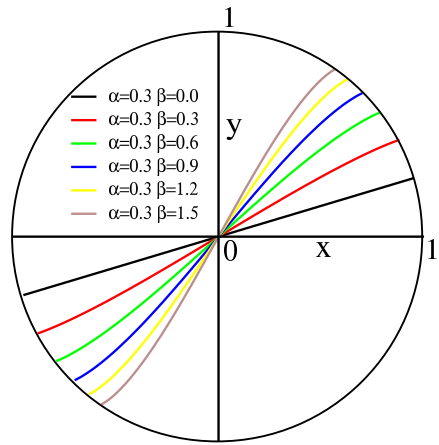}}
\caption{(Color online) Numerical solutions for $y_{surf}$ for $R=1$, $\mu_i = \alpha = 0.3$
  and various values of $\beta = \mu_w/L \, (d \, \Omega^2 / g)^{1/4} $. For $\beta=0$, the influence of the end plates is null and the free surface is flat. When $\beta$ is increased, i.e. increasing influence of the end plates, a curvature appears and the free surface displays an S shape.}
\label{fig3}
\end{center}
\end{figure}


One major criticism can be made regarding the present model: Eq 
(1) was derived for uniform flows and its validity for flows in a
rotating drum is questionable since the flow rate varies
along the flow. Moreover, when the grains have acquired a high
velocity during the flow, they can form an upward tail at the end of
the slope. Therefore our model cannot reproduce this upward tail since the
inertia of the grains is neglected.

\begin{figure}[htbp]
\begin{center}
\resizebox{8.6cm}{!}{\includegraphics*{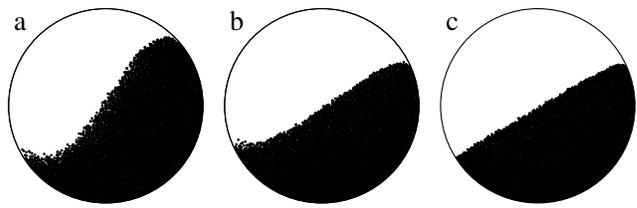}} \caption{Side-views of the 3D-drum. In all cases, $R=40 \,d$, $\mu=0.3$, and $\Omega$=0.2 rad/s. a) $L=10 \,d$, b) $L=50 \,d$, c) $L=300 \,d$. The longer the channel, the flatter the surface. This result gives further support to the idea that the end
plates are responsible for the S shape of the free surface.}\label{fig4}
\end{center}
\end{figure}

The model indicates that there are different ways of changing the
influence of the end plates. One way is to reduce their frictional
properties as demonstrated by Fig. 2, and another way is to increase
the length of the drum. Indeed, when $L$ increases, the second term in
Eq. (6) vanishes and the free surface should tend toward a flat plane. This is demonstrated by Fig. 4,
 which shows three runs with increasing drum length while keeping the radius, filling ratio, frictional 
properties, and rotation speed constant. 
In a narrow drum [Fig. 4a, $L=10 \,d$] the pile displays a well-marked S shape. 
In a longer drum [Fig 4b, $L=50 \,d$] the free surface flattens although a curvature is still visible. 
When the length is further increased [Fig 4c, $L=300 \,d$] the system relaxes to its ground state consisting of a flat surface. This fact gives further support to our model. Once again, since the radius and the rotation speed are identical for all three runs, $Fr$ is identical as well, which shows that in this case, the S shape originates neither from the centrifugal force nor the feeding inertia.

Let us now discuss the scaling laws of the system. First, note that
Eq. (6) can be made dimensionless using the reduced variables
$\tilde{x}=x/R$ and $\tilde{y}=y_{surf}/R$
\begin{equation}
\left\{
\begin{array}{rcl}
\displaystyle \frac{\partial \tilde{y}}{\partial \tilde{x}} &=&  \displaystyle
\mu_i + \mu_w \, \Lambda \, \sqrt{1-\tilde{x}^2-\tilde{y}^2}\\[2mm]
	\Lambda &=& \displaystyle \left(\frac{d \, \Omega^2}{g}\right)^{1/4} \,  \frac{R}{L}
\label{equ7}
\end{array}
\right.
.
\end{equation}


Equation (7) states that the free surface has a ground state consisting of a plane of slope $\mu_i$ from which it deviates when the end plates play an important role.
Moreover, the shape of the free surface in dimensionless units depends only on the value of $\Lambda$ which therefore contains all the scaling laws of the system.
Among them, if the radius $R$ is varied while keeping $d$, $g$, and $L$ constant, the rotation speed should scale as the inverse of $R^2$: $\Omega \propto 1/R^2$. Similarly, one can find the following scaling laws linking the rotation speed to the drum length and particle diameter: $\Omega \propto L^2$ and $\Omega \propto 1/\sqrt{d}$.
In particular, if the radius of the drum is doubled, the rotation
speed should be reduced by a factor of 4 in order to keep $\Lambda$
constant and obtain identical shapes. Similarly, if the length of the drum is doubled, the
rotation speed has to increase by a factor 4. These two examples are
illustrated on Fig. 5 which shows rescaled plots of the free surface
for various values of $R$, $L$, and $\Omega$ while keeping
$\Lambda$ constant. The different sets of data collapse onto one
unique curve, giving numerical confirmation of the scaling laws
inferred from Eq. (7). Let us mention here that the scaling law
$\Omega \propto 1/R^2$ and $\Omega \propto 1/\sqrt{d}$ are compatible
with experimental observations reported by Felix~\cite{Felix2002}.
One can also notice that $\Lambda^4 = Fr \; d/R \; (R/L)^4$. With
the Froude number and the ratio $d/R$ being small it appears that the
end plates have a significant effect only if the aspect ratio $R/L$ is large.

\begin{figure}[t]
\begin{center}
\resizebox{5.3cm}{!}{\includegraphics*{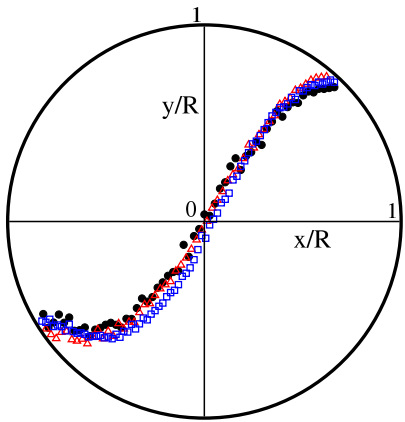}} \caption{
(Color online) Rescaled plots of the position of the free surface for various values of $R$, $L$ and $\Omega$:
Triangles: $R=80d$, $L=10d$, $\Omega= 0.05$rad/s, Squares: $R=80d$, $L=20d$, $\Omega=0.2$rad/s, Circles: $R=40d$, $L=10d$, $\Omega= 0.2$rad/s. In the simulation, the position of the free surface is calculated by identifying the highest grain located at a position $x$. $R$, $L$ and $\Omega$ are varied while keeping $\Lambda$ constant. The data collapse onto one unique curve which validates the proposed scaling laws.
}\label{fig5}
\end{center}
\end{figure}

We have shown that frictional end plates have a major and nontrivial influence on the shape of a granular pile in a rotating drum.
Through numerical simulations we have demonstrated that the S shape disappears when the friction on the end plates vanishes or when the drum is long enough.
Our theoretical model supports the idea that the end plates are responsible for the curvature of the free surface and  shows that the dimensionless number $\Lambda$ (which includes all the relevant parameters: particle size, drum length and radius, rotation speed, and gravity) entirely describes the shape of the pile.
Our theoretical analysis could be improved by including the inertia of the grains. A model similar to that of Khakhar {\it et al.}~\cite{khakhar05} could be adapted by adding a friction term accounting for the effect of the end plates. Such a model will be presented in a future paper.
Finally, we have presented only a few scaling laws regarding the shape of the free surface but many more can be inferred from the expression of $\Lambda$.

The authors would like to thank D.V. Khakhar and A. Caprihan for fruitful discussions.

\end{document}